\documentclass[fleqn,10pt]{wlscirep}
\usepackage[utf8]{inputenc}
\usepackage[T1]{fontenc}
\usepackage{subfigure}
\title{PAD-UFES-20: a skin lesion dataset composed of patient data and clinical images collected from smartphones}

\author[1,4]{Andre G. C. Pacheco}
\author[2,3]{Gustavo R. Lima}
\author[2,3]{Amanda S. Salomão}
\author[4]{Breno A. Krohling}
\author[2,3]{Igor P. Biral}
\author[4]{Gabriel G. de Angelo}
\author[2,3]{Fábio C. R. Alves Jr}
\author[1,4]{José G. M. Esgario}
\author[2,3]{Alana C. Simora}
\author[4]{Pedro B. C. Castro}
\author[4]{Felipe B. Rodrigues}
\author[5, 3, $\dag$,*]{Patricia H. L. Frasson}
\author[1, 4, 6, $\dag$,*]{Renato A. Krohling}
\author[4]{Helder Knidel}
\author[8]{Maria C. S. Santos}
\author[7,3, $\dag$]{Rachel B. do Espírito Santo}
\author[7,3, $\dag$]{Telma L. S. G. Macedo}
\author[7,3, $\dag$]{Tania R. P. Canuto}
\author[3, $\dag$]{Luíz F. S. de Barros}
\affil[1]{Graduate Program in Computer Science, Federal University of Espirito Santo, Vitoria, Brazil}
\affil[2]{Faculty of Medicine, Federal University of Espirito Santo, Vitoria, Brazil}
\affil[3]{Dermatological Assistance Program (PAD), Federal University of Espirito Santo, Vitoria, Brazil}
\affil[4]{Nature Inspired Computing Laboratory, Federal University of Espirito Santo, Vitoria, Brazil}
\affil[5]{Department of Specialized Medicine, Federal University of Espirito Santo, Vitoria, Brazil}
\affil[6]{Production Engineering Department, Federal University of Espirito Santo, Vitoria, Brazil}
\affil[7]{Secretary of Health of the Espírito Santo state, Governor of Espírito Santo state, Vitória, Brazil}
\affil[8]{Pathological Anatomy Unit of the University Hospital Cassiano Antônio Moraes (HUCAM), Federal University of Espirito Santo, Vitoria, Brazil}

\affil[*]{corresponding author(s): R. A. Krohling (rkrohling@inf.ufes.br), P.H.L. Frasson (padufesinstitucional@gmail.com)}
\affil[$\dag$]{these authors jointly supervised this work}

\begin{abstract}

Over the past few years, different computer-aided diagnosis (CAD) systems have been proposed to tackle skin lesion analysis. Most of these systems work only for dermoscopy images since there is a strong lack of public clinical images archive available to design them. To fill this gap, we release a skin lesion benchmark composed of clinical images collected from smartphone devices and a set of patient clinical data containing up to 22 features. The dataset consists of 1,373 patients, 1,641 skin lesions, and 2,298 images for six different diagnostics: three skin diseases and three skin cancers. In total, 58.4\% of the skin lesions are biopsy-proven, including 100\% of the skin cancers. By releasing this benchmark, we aim to aid future research and the development of new tools to assist clinicians to detect skin cancer.

\end{abstract}
\begin{document}

\flushbottom
\maketitle

\thispagestyle{empty}

\section*{Background \& Summary}
Skin lesions are characterized by damage skin cells caused by either genetic or environmental factors \cite{CCA2018}. Most of the skin lesions are benign (non-cancerous) and do not offer a high risk; however, some instances of this disorder may evolve to skin cancer, the most common dysplasia around the world \cite{bray2018global}. In order to diagnose skin cancer, dermatologists screen the skin lesion, assess the patient clinical information, and use their experience to classify the lesion \cite{pacheco2020impact}. Nonetheless, accurate diagnosis is challenging and requires proper training and experience in dermoscopy \cite{kittler2002, sinz2017, tschandl2019comparison}, a noninvasive diagnostic technique that uses optic magnification to permit the visualization of morphologic features that are not visible to the naked eye \cite{argenziano2003dermoscopy}. In this context, the high incidence and the lack of experts have increased the demand for computer-aided diagnosis (CAD) systems for skin cancer.

Over the past few years, different skin lesion datasets composed of dermoscopy images have been fomenting the development of more sophisticated CAD systems. The Atlas of Dermoscopy \cite{argenziano2000interactive} was the first well-known dataset containing over one thousand skin lesions. In 2018, Tschandl, Rosendahl, and Kittler released the HAM10000\cite{tschandl2018ham10000}, a large collection of multi-source dermatoscopic images of common pigmented skin lesions containing over 10 thousand samples. One year later, Combalia et al. presented the BCN20000\cite{combalia2019bcn20000}, a dataset containing around 20 thousand images of skin cancer, including lesions found in hard-to-diagnose locations (nails and mucosa). Together, both HAM10000 and BCN20000 are the majority part of the International Skin Imaging Collaboration (ISIC) archive (\url{https://www.isic-archive.com}), a public repository that plays an important role for both the purposes of clinical training, and for supporting technical research toward automated algorithmic analysis. Since 2016, this organization hosts the ISIC challenge, an open competition that has been boosting automated skin cancer detection algorithms. Recently, the ISIC 2020 challenge (\url{https://challenge2020.isic-archive.com}) was released using 33,126 dermoscopy training images of unique benign and malignant skin lesions from over 2,000 patients from the archive.

Developing CAD systems to detect skin cancer using dermoscopy images is an important and promising task. However, in emerging countries\cite{scheffler2008forecasting} and in remote/rural\cite{feng2018comparison} areas there is a strong lack of experts and dermatoscope available to screen the skin lesions. As a result, this type of CAD system is not feasible for these places. In this context, CAD systems embedded on smartphones may be a low-cost solution to both deal with the lack of dermatoscope and to assist non-expert clinicians to detect skin cancer\cite{Castro20}. However, to develop such system it is necessary clinical images instead of dermoscopy ones.

In order to aid future research and the development of new tools to detect skin cancer, we present the PAD-UFES-20, a dataset composed of clinical images of skin lesions and patient clinical data related to each lesion collected from different smartphone devices. The dataset contains 2,298 samples of six different types of skin lesions, three cancers and three skin diseases. In addition, each image has up to 22 clinical features including the patient's age, skin lesion location, Fitzpatrick skin type, and skin lesion diameter. The dataset was collected from 2018 to 2019 and, to the best of our knowledge, it is the first public archive composed of clinical images and patient data.

\section*{Methods}
The Dermatological and Surgical Assistance Program (in Portuguese: Programa de Assistência Dermatológica e Cirurgica - PAD) at the Federal University of Espírito Santo (UFES-Brazil) is a nonprofit program that provides free skin lesion treatment, in particular, to low-income people who cannot afford private treatment. For historical reasons, the Espírito Santo state has received thousands of immigrants from Europe throughout the 19th century. As Brazil is a tropical country, most of these immigrants and their descendants were/are not adapted to this climate. As a result, there is a high incidence of skin lesions/cancer in this state and the PAD plays a fundamental role to assist these people\cite{frasson2017panorama}. 

In late 2017, the Nature Inspired Computing Laboratory (LABCIN-UFES) and the PAD started a partnership that resulted in the creation of a web-based platform and a multi-platform smartphone application to collect and store patient clinical data and skin lesion images. In this section, we present a brief description of this system and the data collection.

\subsection*{Software description}
The PAD provides full skin lesion treatment, from the screening to the surgical process (if needed), in 11 different countryside cities in Espírito Santo state. Most of these places are rural areas and do not have internet access, which is an important requisite to take into account. In this context, the software infrastructure is composed of three parts: a local web-server, a remote web-server, and a multi-platform smartphone application. Basically, all data is collected using the smartphone application, which locally connects to the local web-server to store the data. After the data collection is done, as soon as the local web-server gets access to the internet, it synchronizes all data with remote one.

Regarding technologies applied to develop the software, the smartphone application was developed using React-Native (\url{https://reactnative.dev}), which is an open-source library, based on Javascript (\url{https://www.javascript.com}), for building user interfaces for both Android and iOS. Both local and remote web-servers were developed using two main frameworks: Angular(\url{https://angular.io}) and SpringBoot  (\url{https://spring.io}). Angular is open-source framework, based on Typescript (\url{https://www.typescriptlang.org}), for developing efficient and sophisticated single-page applications. SpringBoot is also an open-source framework, based on Java (\url{https://www.java.com}), that is used to create a micro service on the server side. The database is managed using MySQL (\url{https://www.mysql.com}). For more information, please, refer to the code availability section.

Beyond store all data in an organized and structured way, the remote web-server offers a friendly interface to clinicians to access the collected data. This is important for three reasons: 1) it is used to train medical students to identify the lesions; 2) it is important to keep tracking patient lesions since evolution is an important feature to pay attention to detect skin cancer\cite{azulay2015}; 3) it helps clinicians with statistics about lesions and patients, which is relevant to understand the behavior of the disease over Espírito Santo state. This software is also open-sourced and the code repository is described in the code availability section.

\subsection*{Data collection}
\begin{figure}[ht]
\centering
\includegraphics[width=\linewidth]{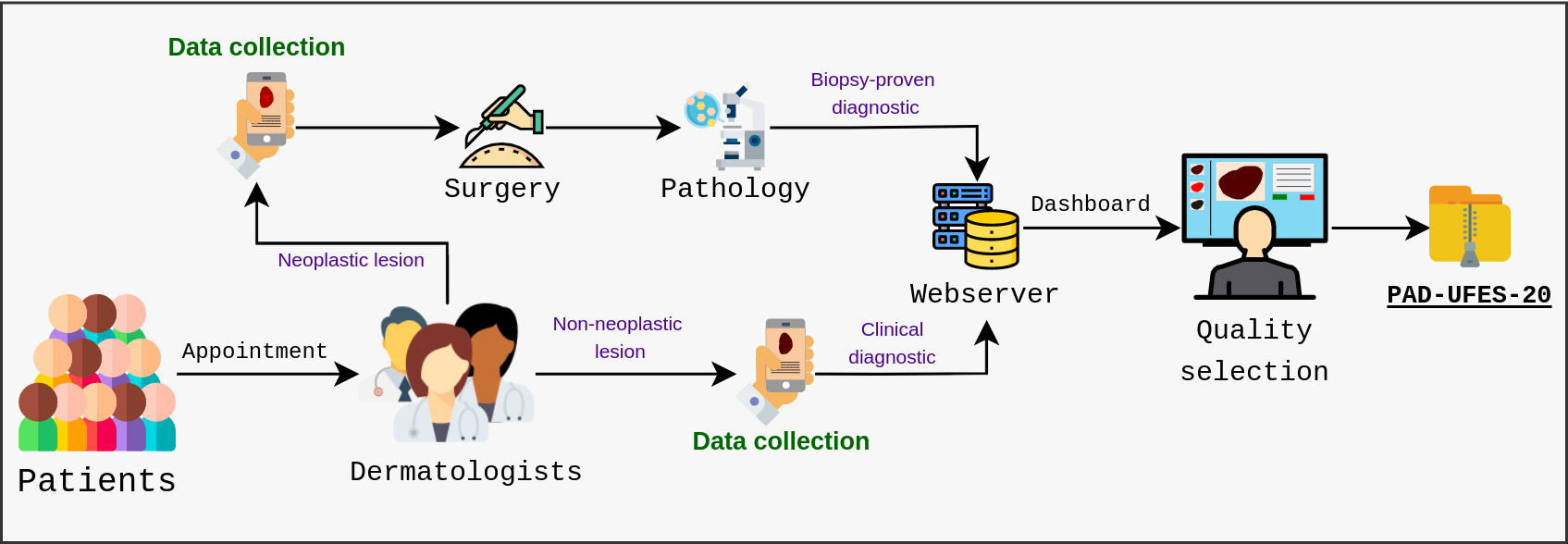}
\caption{Data collection workflow of the PAD-UFES-20 dataset}
\label{fig:data_collection_workflow}
\end{figure}

The data collection workflow is summarized in Fig. \ref{fig:data_collection_workflow}. First of all, the patients have an appointment with a group of up to three senior dermatologists (at least 15 years of experience) that assesses the skin lesion. If the group identifies a neoplasm, the skin lesion is removed through surgical procedure -- performed by medical students under the supervision of two senior plastic surgeons of the PAD -- and sent to the Pathological Anatomy Unit of the University Hospital Cassiano Antônio Moraes (HUCAM) at the UFES to perform histopathology examination. On the other hand, if the group has a consensus that there is no neoplasm, they do not request a biopsy. In both cases, we collect images and clinical data. Later, when the biopsy result is available, it is filled for those lesions in which it was requested. All data is stored in a web-server and the final step is a quality selection to review every single sample that was collected in the previous steps.

The goal of the quality selection step is to review the patient clinical data and remove poor quality images. All data from the appointment are filled by senior medical students and the images are collected using different types of smartphone devices. In addition, the smartphone application allows the user to crop the image to select only the region of interest. As a result, images in different conditions are upload to the server. Thereby, during the quality selection, we delete those images according to the following rules:
\begin{itemize}
    \item The image resolution is very poor and it is not possible to identify the lesion
    
    \item The patient may be identified because of a tattoo, for example
    
    \item The lesion is completely occluded by a hair or ink marking
\end{itemize}

\noindent It is worth noticing that the images present in the dataset have different resolutions, sizes, and lighting conditions. Essentially, an application to detect skin cancer using clinical images needs to deal with such variability. Thus, it aims to simulate the real world. To conclude, all images are raw, i.e., we do not apply any image processing to enhance visualization. 

Regarding the patient clinical data, we review all samples to correct typos and information that are clearly wrong, for example,birth dates before 1900 or lesions' diameter that are way bigger than it looks on the image according to visual inspection. For these cases, we re-checked the physical files in order to fix the information. If it is not possible to fix it, we remove the wrong information from the clinical data and it becomes missing data. Lastly, we translate all clinical data from Brazilian-Portuguese to English.

\subsection*{Data selection}

The dataset was collected during 2018 and 2019. In total, there are over 50 types of skin lesions that were collected during this period. However, most of them are rare and contain only a few samples. For this reason, we selected the seven most common skin lesions diagnosed at PAD, which are: Basal Cell Carcinoma (BCC), Squamous Cell Carcinoma (SCC), Actinic Keratosis (ACK), Seborrheic Keratosis (SEK), Bowen’s disease (BOD), Melanoma (MEL), and Nevus (NEV). As the Bowen’s disease is considered SCC in situ \cite{wolff2017}, we clustered them together, which results in six skin lesions in the dataset, three skin cancers (BCC, MEL, and SCC) and three skin disease (ACK, NEV, and SEK). The number of samples of each skin lesion is presented in Table \ref{tab:freq_diag}. It is important to note that all samples diagnosed as skin cancer presented in this dataset are biopsied proved. For those cases in which the pathology yielded a biopsy that is inconclusive, we removed the sample from the dataset. In Table \ref{tab:freq_diag} is also described the percentage of samples diagnosed as ACK, NEV, and SEK that were biopsied proved. As we described in the next section, we provide this information in the metadata.

To conclude, there are approximately 120 different anatomical regions used by the PAD's dermatologists and pathologists. We clustered these regions in 15 macro-regions that are more frequent and have more potential to raise a skin lesion, they are: face, scalp, nose, lips, ears, neck, chest, abdomen, back, arm, forearm, hand, thigh, shin, and foot. As skin lesions have preferences for some regions of the body \cite{wolff2017, azulay2015}, it is an important feature to consider.

\section*{Data Records}
All samples within PAD-UFES-20 represents a skin lesion of a patient that is composed of an image and a set of metadata. A patient may have one or more skin lesions and a skin lesion may have one or more images. In total, there are 1,373 patients, 1,641 skin lesions, and 2,298 images present in the dataset. Although the number of total samples is not as high as the HAM10000 or BCN20000, we would like to highlight that this is the first step towards building a public dataset of this type. For comparative purposes, the ISIC archive started in 2016\cite{marchetti2018results} containing only 1,279 samples and today it is over 40 thousand. Moreover, we aim to update the dataset every two years by including more samples and more skin lesions. All data record is available on Mendeley Data\cite{padufes20}.

\subsection*{Images}
\begin{figure}[ht]
    \centering
  \subfigure[BCC\label{fig:samples-bcc}]{
       \includegraphics[width=0.22\linewidth]{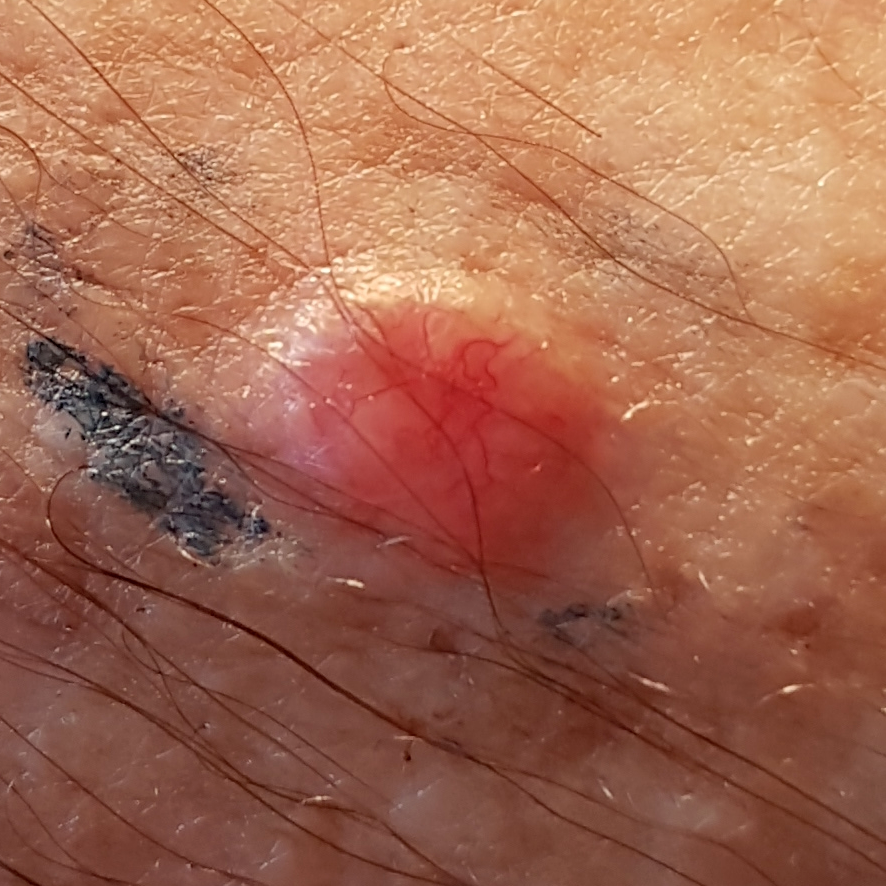}}
    \quad
  \subfigure[SCC\label{fig:samples-scc}]{%
       \includegraphics[width=0.22\linewidth]{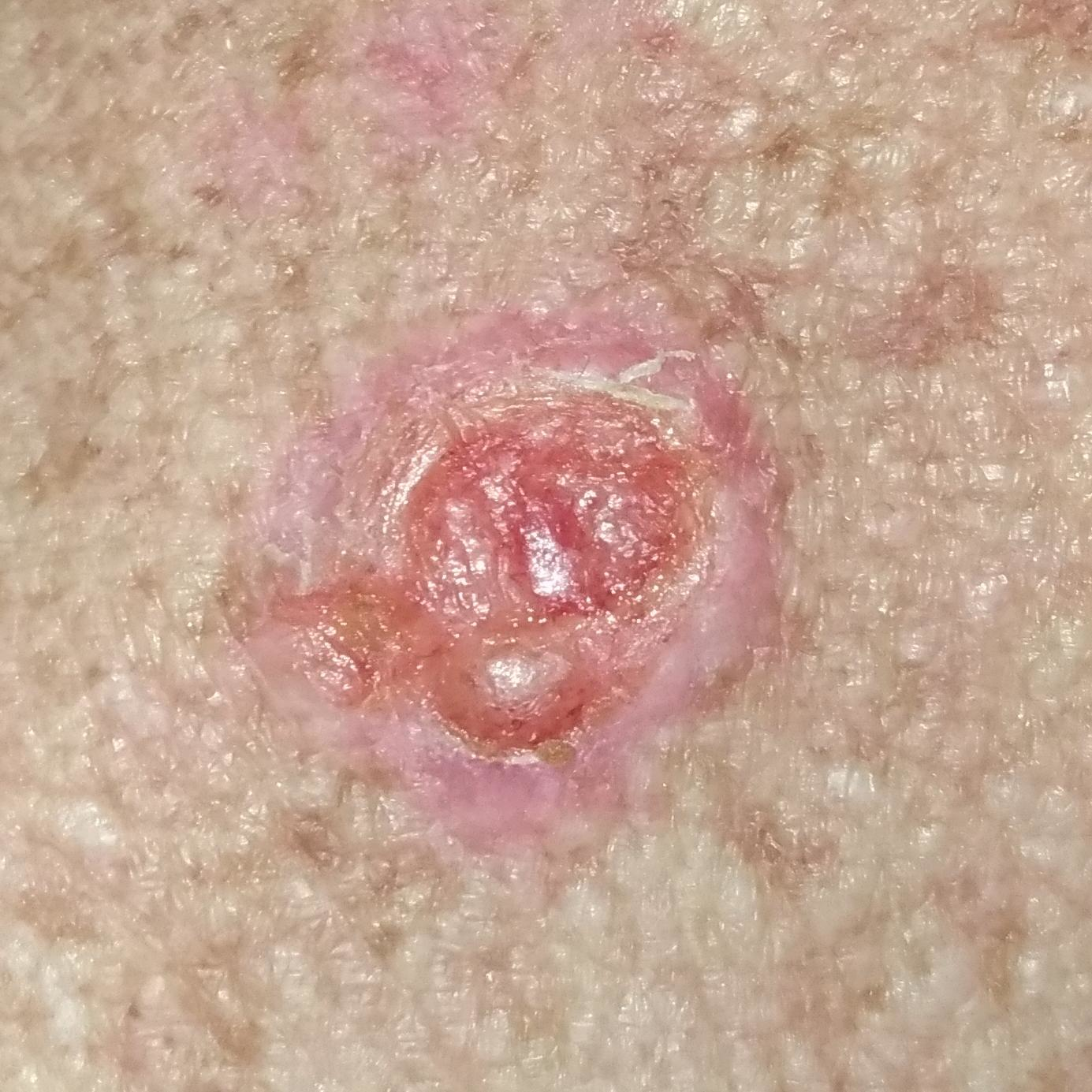}}
    \quad
  \subfigure[ACK\label{fig:samples-ack}]{%
        \includegraphics[width=0.22\linewidth]{src/img/samples/ACK.png}}
    
  \subfigure[MEL\label{fig:samples-mel}]{%
        \includegraphics[width=0.22\linewidth]{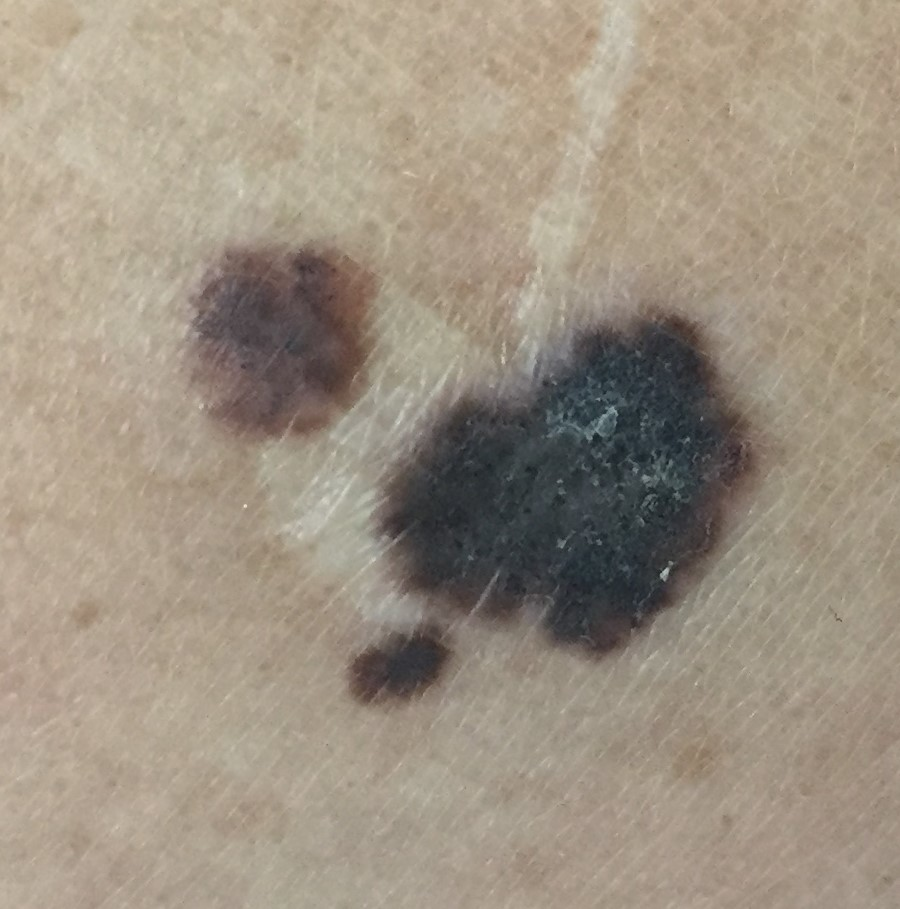}}
    \quad
   \subfigure[NEV\label{fig:samples-nev}]{%
        \includegraphics[width=0.22\linewidth]{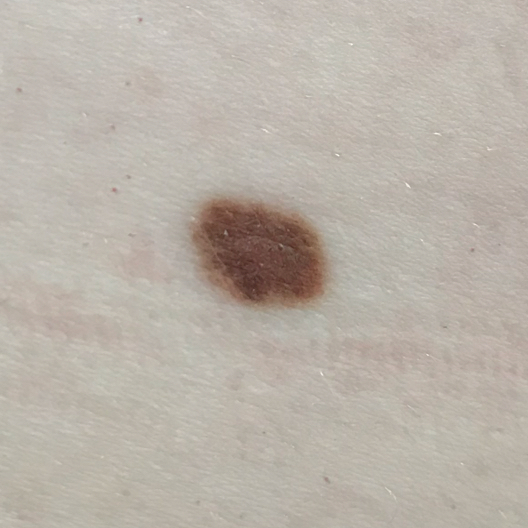}}
    \quad
    \subfigure[SEK\label{fig:samples:f}]{%
        \includegraphics[width=0.22\linewidth]{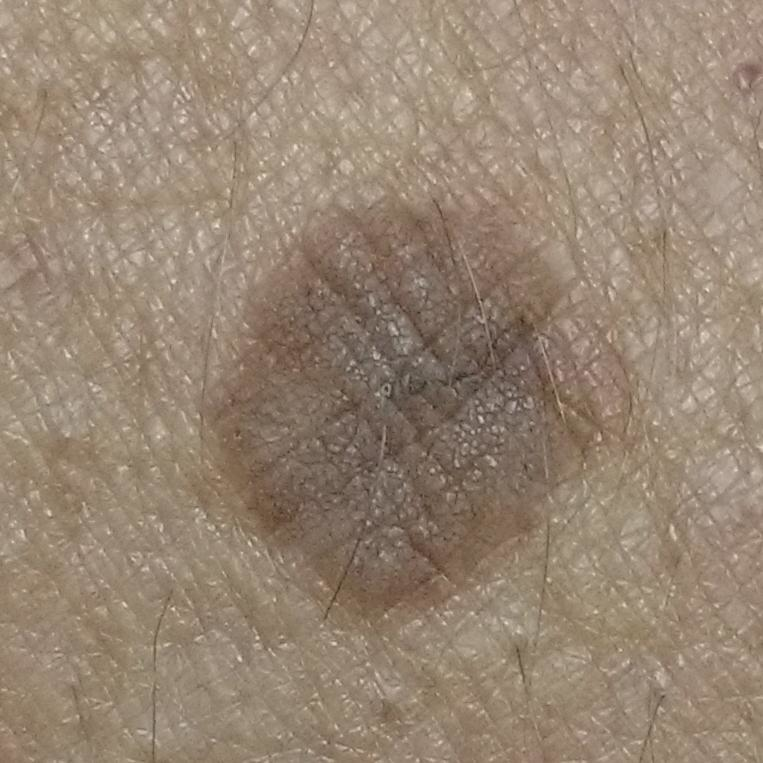}}
        
  \caption{\label{fig:samples}Samples of each type of skin lesion present in PAD-UFES-20 dataset. SCC, BCC, and MEL are skin cancers and NEV, SEK and ACK are skin diseases.} 
\end{figure}

As previously mentioned, the images present in the dataset have different sizes because they are collected using different smartphone devices. All images are available in \texttt{.png} format. In Fig. \ref{fig:samples} is illustrated image samples for each skin lesion present in the dataset.

\subsection*{Meta-data}
The metadata associated with each skin lesion is composed of 26 features. All features are available in a CSV document in which each line represents a skin lesion and each column a metadata feature. We describe all of them in the following:

\begin{itemize}
    \item \textbf{patient\_id}: a string representing the patient ID -- example: \texttt{PAT\_1234}.
    
    \item \textbf{lesion\_id}: a string representing the lesion ID -- example: \texttt{123}.
    
    \item \textbf{img\_id}: a string representing the image ID, which is a composition of the patient ID, lesion ID, and a random number -- example: \texttt{PAT\_1234\_123\_000}.

    \item \textbf{smoke}: a boolean to map if the patient smokes cigarettes.
    
    \item \textbf{drink}: a boolean to map if the patient consumes alcoholic beverages.
    
    \item \textbf{background\_father} and \textbf{background\_mother}: a string representing the country in which the patient's father and mother descends. Note: many patients descend from Pomerania, a region between Poland and Germany. Although it is not a country, we decided to keep the nomenclature, since they identify themselves as Pomeranians descendants.
    
    \item \textbf{age}: an integer representing the patient's age.
    
    \item \textbf{pesticide}: a boolean to map if the patient uses pesticides.
    
    \item \textbf{gender}: a string representing the patient's gender.
    
    \item \textbf{skin\_cancer\_history}: a boolean to map if the patient or someone in their family has had skin cancer in the past.
    
    \item \textbf{cancer\_history}: a boolean to map if the patient or someone in their family has had any type of cancer in the past.
    
    a boolean to map if the patient or anyone in his family has had skin cancer in the past(?)
    
    \item \textbf{has\_piped\_water}: a boolean to map if the patient has access to piped water in their home.
    
    \item \textbf{has\_sewage\_system}: a boolean to map if the patient has access to a sewage system in their home.
    
    \item \textbf{fitspatrick}: a integer representing the Fitspatrick skin type\cite{wolff2017}.
    
    \item \textbf{region}: a string representing one of the 15 macro-regions previously described.
    
    \item \textbf{diameter\_1} and \textbf{diameter\_2}: a float representing the skin lesions' horizontal and vertical diameters. 
    
    \item \textbf{diagnostic}: a string representing the skin lesion diagnostic (ACK, BCC, MEL, NEV, SCC, or SEK).
    
    \item \textbf{itch}: a boolean to map if the skin lesion itches.
    
    \item \textbf{grew}: a boolean to map if the skin lesion has recently grown.
    
    \item \textbf{hurt}: a boolean to map if the skin lesion hurts.
    
    \item \textbf{changed}: a boolean to map if the skin lesion has recently changed.
    
    \item \textbf{bleed}: a boolean to map if the skin lesion has bled.
    
    \item \textbf{elevation}: a boolean to map if the skin lesion has an elevation.
    
    \item \textbf{biopsed}: a boolean to map if the diagnostic comes from a clinical consensus or biopsy.

\end{itemize}

It is important to note that some features may be missing for some lesions. In brief, \texttt{patient\_id}, \texttt{lesion\_id}, \texttt{img\_id}, \texttt{age}, \texttt{region}, and \texttt{biopsed} are always present. The remaining ones depend on the patient's answers during the appointment. Missing values are left blank in the CSV document. When a patient does not know the answer for some question -- for example, he/she does not know their father's background -- we fill the feature as UNK (unknown).

\section*{Technical Validation}
As previously stated, the dataset is composed of clinical and biopsied diagnostics. For NEV, SEK, and ACK, only the clinical diagnosis is performed according to the PAD’s dermatologists consensus during the patient’s appointment. As these skin lesions are benign, it is not necessary to perform a skin biopsy. On the other hand, for skin lesions that dermatologists suspect malignancy, a biopsy is requested for diagnostic confirmation. In this dataset, all BCC, SCC, and MEL are biopsy-proven.
The histopathology procedure involves the following steps: the collection of a skin fragment, tissue fixation in formaldehyde (at a concentration of 10\%), macroscopic analysis of the skin fragment, histological processing, producing the microscope slides, and a microscopic study with diagnosis' formulation and interpretation \cite{werner2009biopsia}. As described in Table \ref{tab:freq_diag}, 58.4\% of the skin lesions in PAD-UFES-20 dataset are biopsy-proven. This number is compatible with other skin lesion datasets described in literature, for instance, the HAM1000\cite{tschandl2018ham10000} has 53.3\% of biopsy-proven samples.

Regarding the metadata features, they were collected according to the anamnesis of a patient, which beyond the skin lesion screening, dermatologists also consider the anatomical region, diameter, ulceration, itching, bleeding, among others characteristics of the skin lesion \cite{wolff2017, azulay2015}. In addition, risk factors are also taken into accounts such as exposure to chemicals, cancer history, and the type of skin \cite{duarte2018risk}. Combining patient clinical data and skin lesion images has been proved to improve the efficacy of CAD systems for skin cancer detection \cite{kharazmi2018, pacheco2020impact}.

\section*{Additional Notes}
There are two main characteristics that differ the PAD-UFES-20 dataset from other skin lesion datasets available on literature: the clinical images collected from smartphone devices and the set of patient clinical data. Beyond educational purposes, this dataset aims to aid the development of CAD systems embedded in smartphones, in particular, to assist clinicians/non-experts to handle skin lesions in remote places. In addition, the dataset contains samples of six of the most commonly known skin lesions \cite{wolff2017}, including pigmented and non-pigmented ones. 

\begin{figure}[ht]
    \centering
  \subfigure[Age distribution\label{fig:age_dist}]{
       \includegraphics[width=0.45\linewidth]{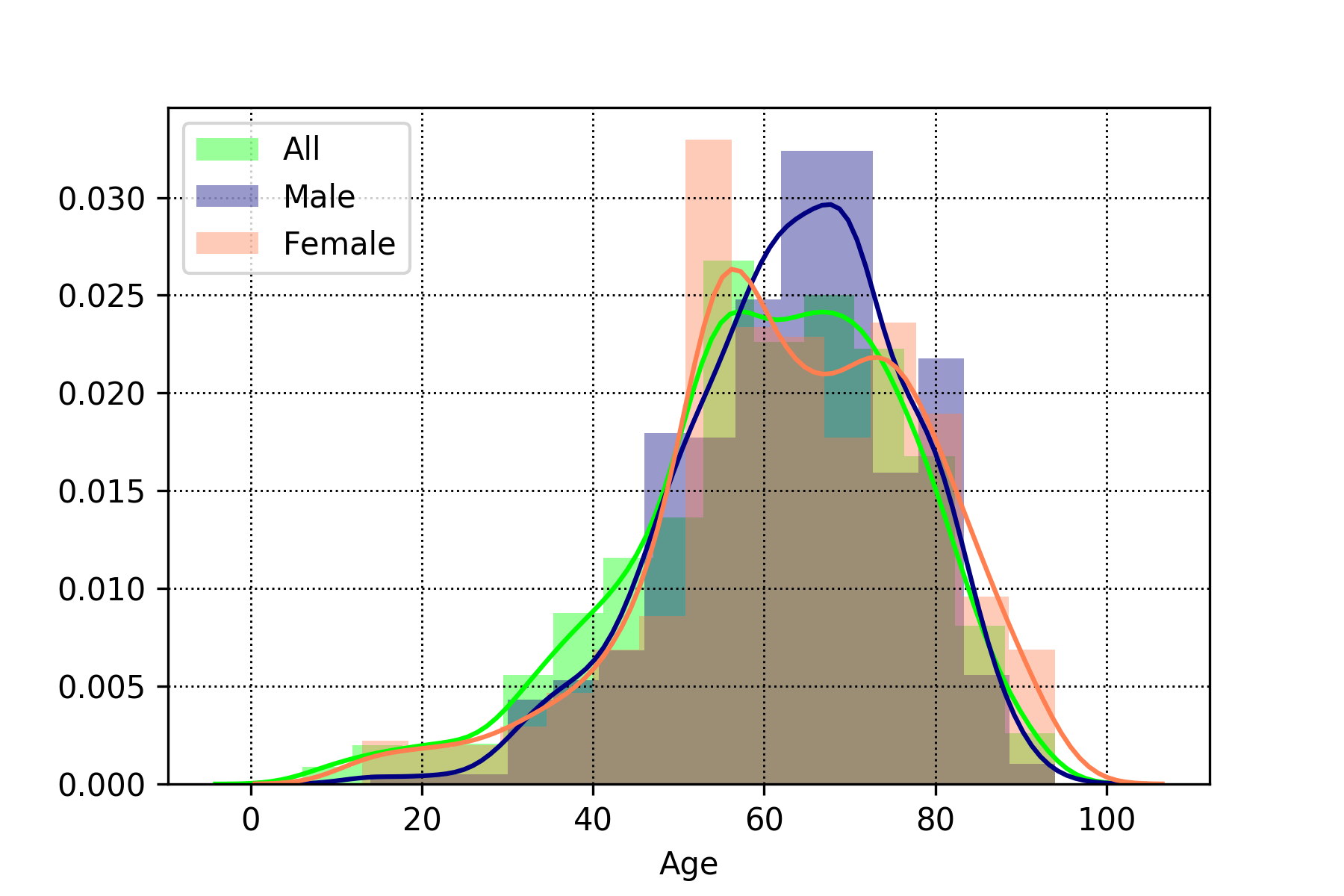}}
    \quad
  \subfigure[Age boxplots\label{fig:age_bp}]{%
       \includegraphics[width=0.45\linewidth]{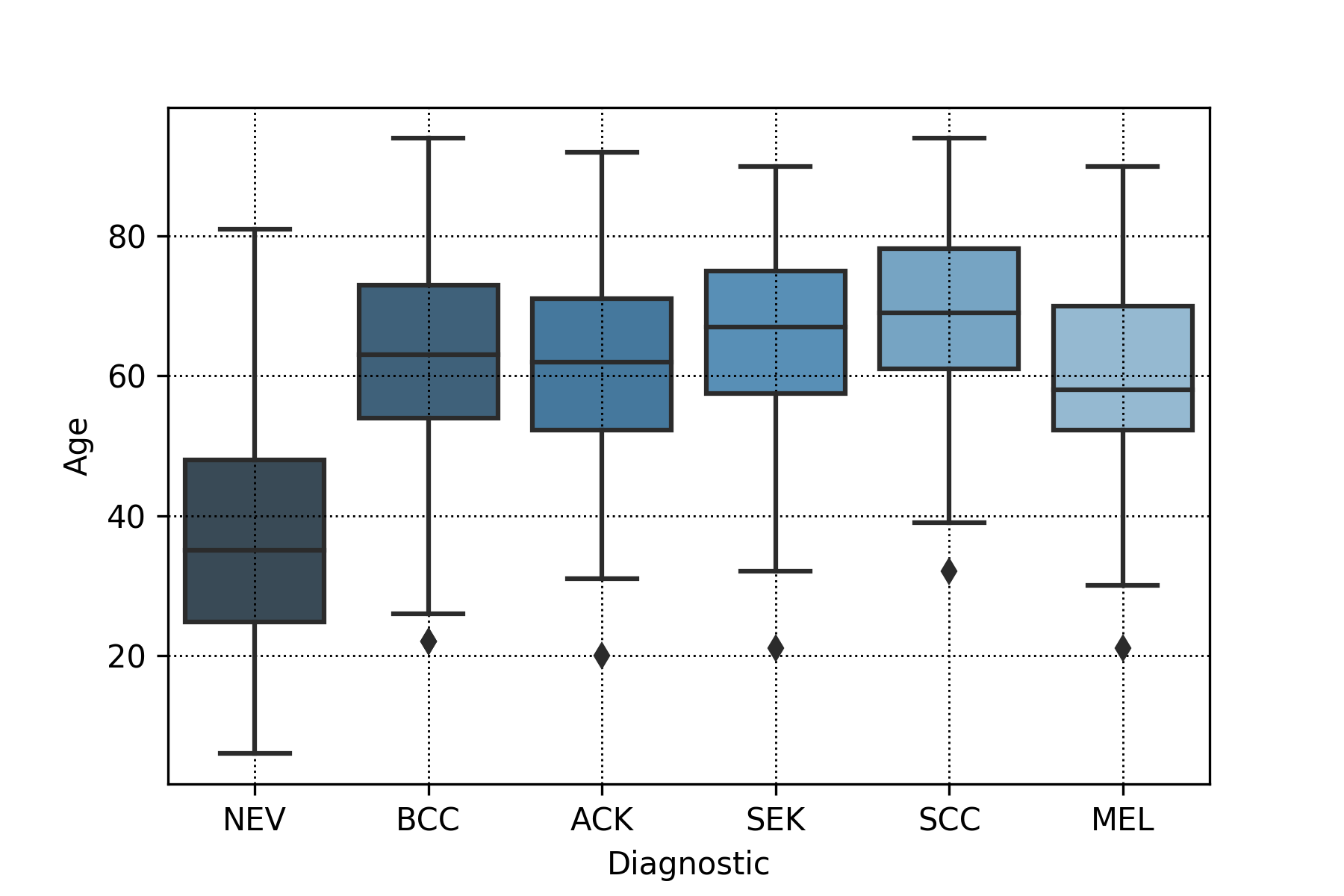}}
  \caption{\label{fig:age} The patients age distribution according to gender and the age boxplots for each diagnostic} 
\end{figure}

All data are collected in 11 different cities in Espírito Santo state, Brazil. Most of the patients are European immigrants descendants and are or have been farm workers with many hours of sun exposure per day. In addition, the patient average age is approximately 60 years old, but it may vary according to the diagnostic -- see Fig. \ref{fig:age}. Thus, it is important to note this data represents a specific population in a particular region in Brazil.

Another important aspect is the raw image. We do not apply any image processing algorithm into the collected images. A suggestion to enhance the images would be using color constancy algorithms, which have been proved to be helpful for automated skin cancer detection \cite{barata2014, pacheco2020impact}. To conclude, as we may note in Table \ref{tab:freq_diag}, the dataset is imbalanced, in particular, for melanoma, the deadliest case of skin cancer. Unfortunately, imbalanced datasets are quite common for skin lesion datasets. For instance,  
in HAM10000 and in BCN20000, we may find an imbalance among the diagnostic labels of approximately 58:1 and 40:1 respectively. For automated skin cancer detection, common algorithms to deal with this issue are oversampling and weighted loss functions\cite{pacheco2020impact}.

\section*{Code availability}
The code of the software developed to collect data, the script to validate the meta-data, and a brief exploratory data analysis is available on \url{https://github.com/labcin-ufes/PAD-UFES-20}

\bibliography{sample}

\section*{Acknowledgments}
This study was financed in part by the Coordena\c{c}\~{a}o de Aperfei\c{c}oamento de Pessoal de N\'{i}vel Superior - Brasil (CAPES) - Finance Code 001; the Conselho Nacional de Desenvolvimento Cient\'{i}fico e Tecn\'{o}logico (CNPq) - grant n.309729/2018-1 - and the Funda\c{c}\~{a}o de Amparo {a} Pesquisa e Inova\c{c}\~{a}o do Esp\'{i}rito Santo (FAPES) - grant n. 575/2018. We also thank the support of the Secretary of Health of the Espírito Santo state (SESA), the municipal administration of the 11 cities in which PAD takes place, and the Lutheran church of the Espírito Santo state. Lastly, we acknowledge the work of Prof. Carlos Cley and Prof. Luiz F. S. de Barros who founded the PAD in 1987.

% \section*{Author contributions statement}
% All authors contributed to data collection. R.A.K. and H.K. conceived the idea of this dataset. A.G.C.P. led software technical development and, along with G.R.L., A.S.S, I.P.B., A.C.S., and F.C.R.A., coordinated the data collection. B.A.K., J.G.M.E., P.B.C.C., and F.B.R., worked on software development. P.H.L.F. and L.F.S.B. are the coordinators of the PAD and the plastic surgeons. M.C.S.S. is responsible for histopathology examination and description.  R.B.E.S., T.L.S.G., and T.R.P.C. are the dermatologists of the PAD and, along with P.H.L.F., R.A.K, and L.F.S.B., they are the supervisors of this work. A.G.C.P. wrote the initial draft of the manuscript with input and revisions from R.A.K., P.H.L.F., R.B.E.S, B.A.K., G.R.L., and A.S.S.

\section*{Competing interests} 
The authors declare no competing interests.

\section*{Declaration of consent} 
The dataset was collected along with the Dermatological and Surgical Assistance Program (PAD) of the Federal University of Espírito Santo. The program is managed by the Department of Specialized Medicine and was approved by the university ethics committee (nº 500002/478) and the Brazilian government through Plataforma Brasil (nº 4.007.097), the Brazilian agency responsible for research involving human beings. In addition, all data is collected under patient consent and the patient’s privacy is completely preserved. 

\section*{Figures \& Tables}

\begin{table}[ht]
\centering
\begin{tabular}{|c|c|c|}
\hline
Diagnostic                    & Nº of samples& \% biopsied \\ \hline
Actinic Keratosis (ACK)       & 730          & 24.4\%       \\ \hline
Basal Cell Carcinoma (BCC)    & 845          & 100\%        \\ \hline
Melanoma (MEL)                & 52           & 100\%        \\ \hline
Nevus (NEV)                   & 244          & 24.6\%       \\ \hline
Squamous Cell Carcinoma (SCC) & 192          & 100\%        \\ \hline
Seborrheic Keratosis (SEK)    & 235          & 6.4\%        \\ \hline
\textbf{Total}                & \textbf{2298}         & \textbf{58.4}\%       \\ \hline
\end{tabular}
\caption{\label{tab:freq_diag}The number of samples for each type of skin disorder present in the PAD-UFES-20 dataset}
\end{table}

\end{document}